\documentclass{article}
\usepackage{spconf,amsmath,graphicx}
\usepackage{dsfont}
\usepackage{algorithm}
\usepackage{algpseudocode}
\usepackage{multirow}
\usepackage[table,xcdraw]{xcolor}
\title{Quanta Diffusion}
\name{Prateek Chennuri$^{\dagger}$ \qquad Dongdong Fu$^{\star}$ \qquad Stanley H. Chan$^{\dagger}$}
  
  \address{$^{\star}$Dolby Laboratories, Sunnyvale CA 94085, USA \\
      $^{\dagger}$Purdue University, West Lafayette IN 47907, USA}
\begin{document}
\maketitle
\renewcommand{\thefootnote}{}% Remove number
\footnotetext{The work is supported, in part, by the National Science Foundation under
the grants 2133032, 2030570, and Angular Encoded Imaging.}
\begin{figure*}[ht]
    \centering
    \includegraphics[width=0.95\linewidth, trim=15mm 0 15mm 0, clip]{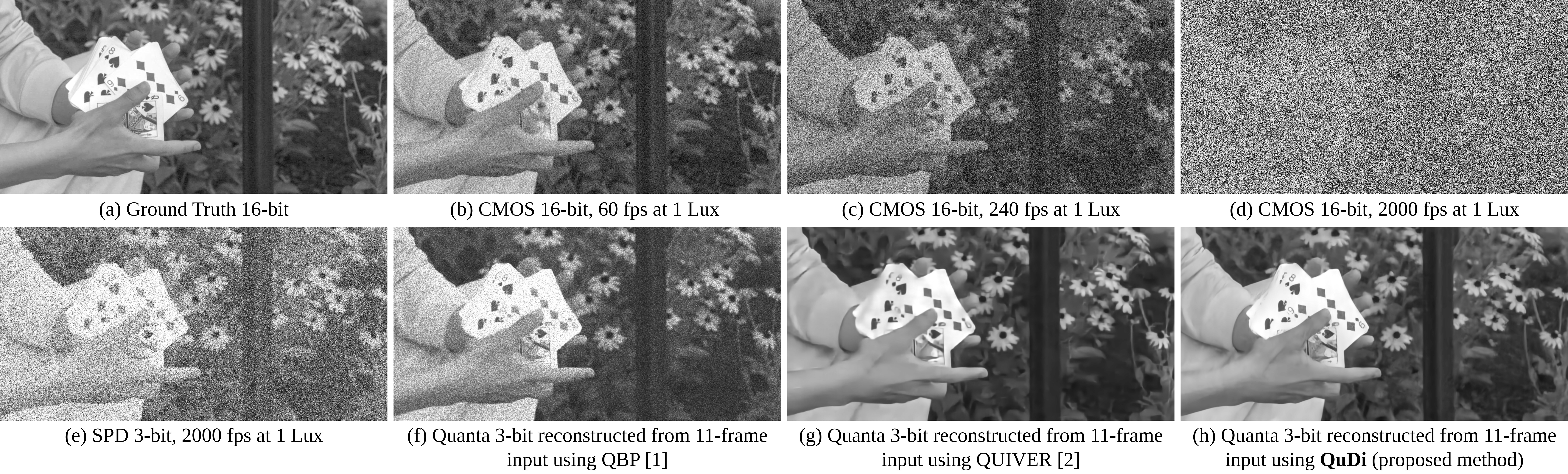}
    % \vspace{-2.5ex}
    \caption{\textbf{Project Goal}. (a) Blur-free video frame of a card-based magic trick. (b)-(d) CMOS image sensor simulations using realistic sensor parameters. The strong shot noise and read noise ($5.1\;\text{e}^-$/pix) of CMOS sensor make the signal acquisition difficult. (e) With low read noise ($0.2\;\text{e}^-$/pix), low-bit Single-Photon Detectors (SPDs) capture valuable information. (f) Existing state-of-the-art classical method, QBP~\cite{ma2020quanta} cannot handle strong motion and noise. (g) Existing state-of-the-art deep-learning based method, QUIVER~\cite{chennuri2024quiver} produces over-smoothened results. (h) The proposed algorithm, QuDi, produces high perceptual-quality outputs. \textit{Best viewed in zoom.}
    }
    \label{fig:problem_scope}
\end{figure*}
\begin{abstract}
We present Quanta Diffusion (QuDi), a powerful generative video reconstruction method for single-photon imaging. QuDI is an algorithm supporting the latest Quanta Image Sensors (QIS) and Single Photon Avalanche Diodes (SPADs) for extremely low-light imaging conditions. Compared to existing methods, QuDi overcomes the difficulties of simultaneously managing the motion and the strong shot noise. The core innovation of QuDi is to inject a physics-based forward model into the diffusion algorithm, while keeping the motion estimation in the loop. QuDi demonstrates an average of $2.4$ dB PSNR improvement over the best existing methods.
\end{abstract}

\section{Introduction}
\label{sec:Intro}
The invention of Single Photon Detectors (SPDs)~\cite{fossum2011quanta} has fundamentally revolutionized the landscape of computational imaging especially in applications like low-light imaging~\cite{chi2020dynamic, chan2016images}, high-speed videography~\cite{ma2020quanta}, time-of-flight sensing~\cite{weerasooriya2025joint}, and 3D imaging~\cite{lindellSinglephoton3DImaging2018}. The critical advantage single photon detectors possess over the conventional CMOS sensors is their highly-accurate photon counting ability along with their ultra-high temporal resolution. Although SPDs have the intrinsic capability to capture useful information in extreme low-light ($\approx 1~\text{Lux}$) and fast motion, high photon-shot noise and dark noise necessitates the use of a post-processing image/video reconstruction algorithm. Therefore, SPDs mostly are accompanied with the question of how do we recover the image/video from the noisy photon count streams while avoiding distortions and motion blur.

To provide a visual representation of the problem's scope, Figure~\ref{fig:problem_scope} depicts a blur-free video of a fast-moving cards. We simulate CMOS Image Sensor (CIS) captures at $60$ fps, $240$ fps, and $2000$ fps using realistic sensor parameters. We see that the
CMOS outputs are either extremely blurred due to fast motion or severely noisy due to sparse photons and inherently high read noise. We also simulate the SPD output as shown in Figure~\ref{fig:problem_scope}. Despite heavy photon shot noise, SPDs record useful information which can be utilized for restoring the image/video. Existing quanta-burst algorithms like QBP~\cite{ma2020quanta} fail to handle noise (in the presence of limited input frames) and motion (in the presence of extensive input frames) simultaneously. While the state-of-the-art QUIVER~\cite{chennuri2024quiver} does handle both noise and motion at the same time with limited input frames, it produces sub-optimal perceptual quality while missing critical high-frequency information as shown in Figure~\ref{fig:problem_scope}. Therefore, with extremely limited input information, it is critical to hallucinate realistic details for enhanced perceptual quality.

In this paper, we propose \textbf{Qu}anta \textbf{Di}ffusion (QuDi), a \underline{physics-based} generative framework conditioned on the quanta data to produce outputs with enhanced perceptual quality. The main contribution(s) of this work can be summarized as follows:
\begin{itemize}
    \item QuDi is the first diffusion-based restoration method proposed for quanta imaging. We explicitly integrate a physics-based probabilistic forward model into the reverse diffusion process, thereby improving the reconstruction quality. 
    \item QuDi has a learnable optical flow estimation module built into the diffusion loop that allows the model to simultaneously handle noise and motion, which are very challenging for previous methods. 
\end{itemize}

\section{Related Work}
\textbf{Image and Video Denoising}. Traditional image and video denoising methods often rely on non-local approaches to identify similar patches across the data~\cite{lebrunNonlocalBayesianImage2013}. Recent advances in deep learning have achieved state-of-the-art performance in denoising tasks~\cite{liUnidirectionalVideoDenoising2022, liangRecurrentVideoRestoration2022a}. However, these methods typically assume simple noise statistics, leading to suboptimal performance in real noisy scenarios. For low-light imaging, burst denoising aligns and merges multiple exposures to enhance quality, though robust alignment remains a challenge.
% \textbf{SPADs and Spike Cameras}. Single Photon Avalanche Diodes (SPADs) enable high-speed imaging in low light with pico-second resolution~\cite{gyongySinglePhotonTrackingHighSpeed2018}. Applications include motion tracking and passive imaging~\cite{ma2020quanta}, though the high temporal resolution imposes bandwidth limitations. Alternative sensors, such as event and spike cameras~\cite{zhaoHighSpeedMotionScene2020}, excel in capturing luminance changes but are unsuitable for extreme low-light conditions due to their thresholding mechanisms~\cite{dong2021spike}.

\textbf{QIS Reconstruction}. Reconstruction of quanta images from Quanta Image Sensors (QIS) presents challenges due to their Poisson-Gaussian noise characteristics. Early methods include optimization-based approaches~\cite{chanEfficientImageReconstruction2014} and transform-based techniques~\cite{chan2016images}. Deep learning has further advanced QIS reconstruction with networks like vision transformers~\cite{wangSinglePhotonCamerasImage2023} and Dual Prior Integrated models~\cite{zhangDualPriorIntegratedImage2023}. However, these approaches struggle with motion, especially in dynamic scenes~\cite{chi2020dynamic}. QUIVER~\cite{chennuri2024quiver}, the current state of the art, achieves superior restoration but suffers from spatio-temporal artifacts when rendered as video.

\textbf{Diffusion Models for Restoration.} Due to its efficacy on mimicing distributions,
diffusion models~\cite{ho2020denoising} have been used for several inverse problems where the output is sampled from the posterior $p_\theta(x|y)$~\cite{kawar2022denoising, tewari2024diffusion}. Among conditional diffusion methods, Tewari et al.~\cite{tewari2024diffusion} is the only method which utilizes a forward model in the diffusion process. However, their forward model is deterministic. In contrast, our work incorporates a probabilistic simulator (Section~\ref{subsec:qis_sim}) that contributes to the random nature of
the diffusion models, thus helping them to produce plausible images conditioned on the input.

\section{Imaging Model and Diffusion Methods}
\subsection{Single Photon Detector Forward Model}
\label{subsec:qis_sim}
We briefly discuss the data acquisition process for a single photon detector in sparse photon conditions. We follow the initial prototype suggested in~\cite{ma2017photon} which was also adopted in prior works such as~\cite{chennuri2024quiver, chi2020dynamic}. 
% We follow the initial prototype suggested in~\cite{ma2017photon} which was also adopted in prior works such as~\cite{chi2020dynamic, chan2016images, chennuri2024quiver}. 

Given the quanta exposure, $\mathbf{I}^{\textbf{GT}}$ that is dependent on the photon flux and the exposure time, the observed signal by the sensor can be modeled as a Poisson-Gaussian random variable where the Poisson represents the photon arrival process and the Gaussian models the read noise. The read-out process involves various sorts of distortions and the Analog-to-Digital Converter (ADC) to represent the overall read-out as integers between $[0, \text{N}]$ where $\text{N} = 2^{\text{Nbits}} - 1$, $\text{Nbits}$ indicating the bit-depth of the image sensor. The final sensor readout, \textbf{Y}, can be mathematically represented as, 
\begin{align}
    \mathbf{Y} \sim \; \text{ADC}_{\left[0, \text{N}\right]} (\text{Poisson}(&\text{QE} \times \mathbf{I}^{\text{GT}} + \theta_{\text{dark}}) \;+ \notag \\ 
    &\underbrace{\text{Gauss}(0, \sigma^2_{\text{read}}\mathbf{I})}_\text{read noise}, \; \text{FWC}).\label{eq: mean_ph_arrival}
\end{align}
where the $\text{QE}$ represents the quantum efficiency, $\theta_{\text{dark}}$ is the dark current, $\sigma^2_{\text{read}}$ indicates the read-noise, and $\text{FWC}$ is the full-well capacity of the image sensor. 

Akin to previous works~\cite{chennuri2024quiver, chi2020dynamic} we assume our sensor is monochromatic and we also utilize monochromatic data for our real experiments. For our sensor prototype, we utilize a Quantum Efficiency (QE) of $0.80$. The dark current ($\theta_{\text{dark}}$) and read noise ($\sigma_{\text{read}}$) are set to $1.6\,\text{e}^-$/pix/frame and $0.2\,\text{e}^-$/pix, respectively. 
\begin{figure*}
    \centering
    \includegraphics[width=0.9\linewidth]{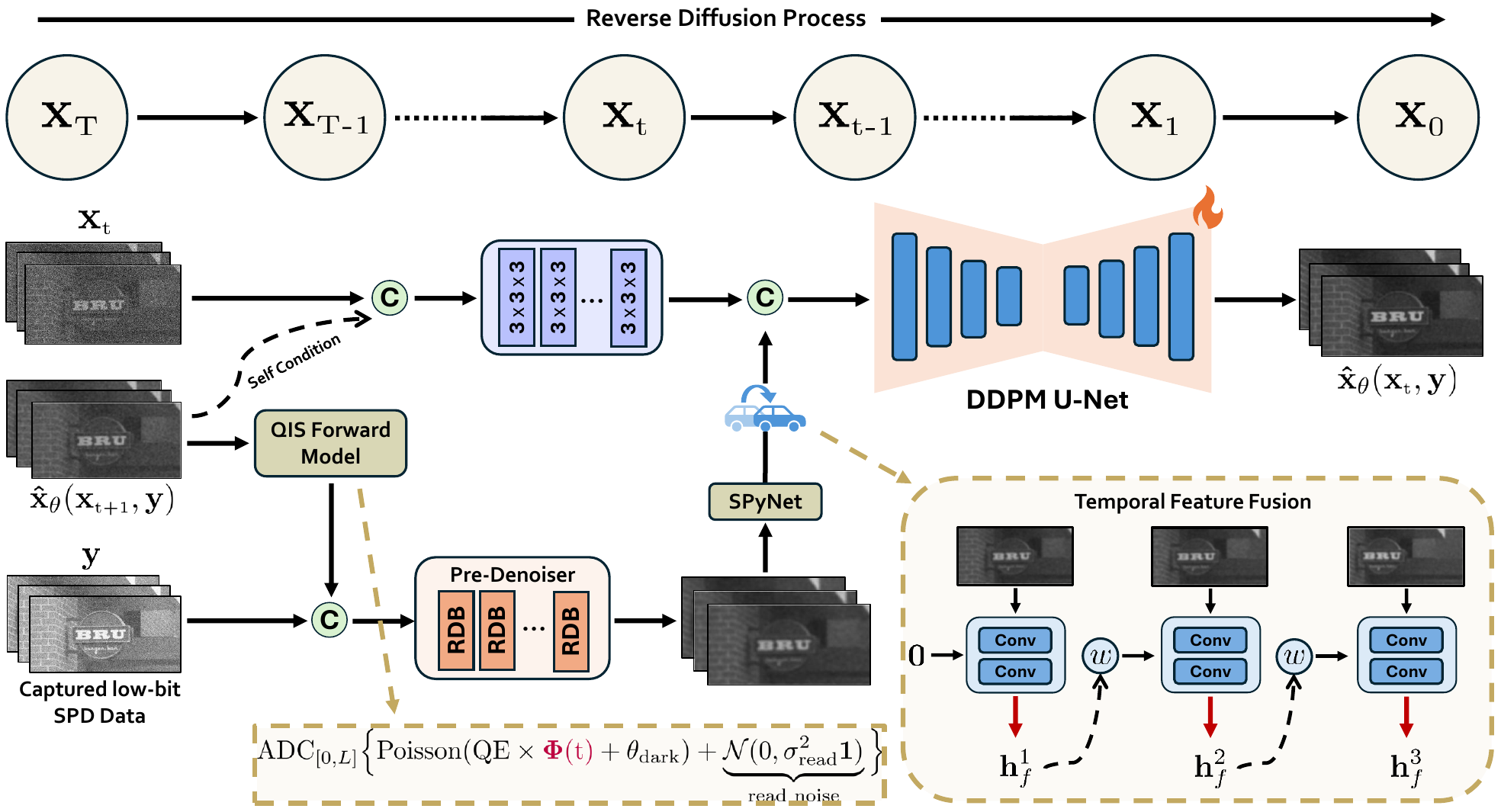}
    % \vspace{-2.5ex}
    \caption{\textbf{Quanta Diffusion Overview.} Inspired by~\cite{tewari2024diffusion}, we utilize the DDPM~\cite{ho2020denoising} U-Net architecture and significantly modify it into a physics-based conditional generative framework specifically designed for quanta restoration. More details available in Section~\ref{sec:qudi}.}
    \label{fig:qudi}
\end{figure*}
\subsection{Diffusion Models}
Diffusion models are a class of generative models that have shown exceptional performance in solving inverse problems compared to traditional CNNs~\cite{kawar2022denoising}. These models learn a data distribution $p_\theta(\mathbf{x})$ to approximate the original distribution $q(\mathbf{x})$. Their Markov chain structure, $\mathbf{x}_\text{T} \rightarrow \mathbf{x}_\text{T-1} \rightarrow \cdots \rightarrow \mathbf{x}_0$, defines the reverse diffusion process:
\begin{equation} 
p_\theta(\mathbf{x}_{0:\text{T}}) = p^\text{(T)}_\theta(\mathbf{x}_\text{T}) \prod_{t=0}^{\text{T}-1} p^\text{(t)}_\theta(\mathbf{x}_\text{t} | \mathbf{x}_\text{t+1}). 
\end{equation}
The forward diffusion process during training is:
\begin{equation} 
q(\mathbf{x}_{1:\text{T}}|\mathbf{x}_0) = q^\text{(T)}(\mathbf{x}_\text{T}|\mathbf{x}_0) \prod_{t=0}^{\text{T}-1} q^\text{(t)}(\mathbf{x}_\text{t} | \mathbf{x}_\text{t+1}, \mathbf{x}_0), \end{equation}
which leads to the Evidence Lower BOund (ELBO). Assuming Gaussian processes for both forward and reverse diffusion, the ELBO simplifies to:
\begin{equation} 
\sum_{\text{t} = 1}^\text{T} \alpha_\text{t} \mathds{E}_{(\mathbf{x}_0, \mathbf{x}_t)\sim q(\mathbf{x}_0)q(\mathbf{x}_\text{t}|\mathbf{x}_0)}\left[\parallel \mathbf{x}_0 - f_\theta^\text{(t)}(\mathbf{x}_\text{t})\parallel_2^2\right], \end{equation}
where $f_\theta^\text{(t)}$ is a neural network trained to denoise $\mathbf{x}_\text{t}$, and $\alpha_{1:\text{T}}$ are coefficients dependent on $q(\mathbf{x}_{1:\text{T}}|\mathbf{x}_0)$. 

\section{Quanta Diffusion}
\label{sec:qudi}
As noted in Section~\ref{sec:Intro}, current quanta restoration methods~\cite{chennuri2024quiver, ma2020quanta} perform well under favorable conditions like ambient light, slow motion, and extensive frame inputs. However, they struggle in extreme low light, fast motion, and limited frame scenarios due to severe photon shot noise in SPD frames. Increasing frame inputs could mitigate noise but introduces errors in long-range flow estimation under fast motion and noise.

It is clear that in extreme conditions, one would need to utilize limited input quanta frames, yet restore details that match the user desired quality. One way is to hallucinate (or generate) realistic information conditioned on the quanta frames. Diffusion Models~\cite{kawar2022denoising, dhariwal2021diffusion}, being the forefront of generative AI could be utilized to synthesize high quality real-world-like details, thus enabling the users to perceive the environment in extreme low-light conditions. 

Building on the discussed challenges and solutions, we propose a \underline{multi-frame} physics-based generative framework, \textbf{Qu}anta \textbf{Di}ffusion (\textbf{QuDi}), illustrated in Figure~\ref{fig:qudi}. Inspired by QUIVER \cite{chennuri2024quiver} and classical quanta burst methods~\cite{ma2020quanta, gyongySinglePhotonTrackingHighSpeed2018}, QuDi leverages the core principles of these approaches. Figure~\ref{fig:qudi} outlines a single step of the reverse diffusion process. The conditional reverse diffusion begins with Gaussian noise ($\mathbf{x}_{\text{T}} \sim \mathcal{N}(0,\mathbf{I})$) and noisy quanta frames ($\mathbf{y} \sim \text{Eq.}$~\ref{eq: mean_ph_arrival} $\in \mathds{R}^{\text{H}\times\text{W}\times\text{N}}$, where $\text{N}$ is the number of frames) to generate a restored noiseless frame $\mathbf{\hat{x}}_\theta(\mathbf{x}_\text{T}, \mathbf{y})$. At any arbitrary iteration, the QuDi architecture can be divided into $4$ main stages as described below:

\textbf{Predenoising to increase SNR.} Existing classical quanta burst methods like~\cite{ma2020quanta} utilize naive averaging method to increase the Signal-to-Noise Ratio (SNR). Although naive averaging proves to be the best method when the motion between the frames is negligible, it produces significant motion blur in the case of fast moving scenes, in turn affecting the flow estimation accuracy. Following QUIVER's analogy~\cite{chennuri2024quiver}, we design a computationally efficient single image denoiser as it helps remove dependency on inter-frame motion. As shown in Figure~\ref{fig:qudi}, though the initial denoised frames look heavily oversmoothened, it provides extensive details on the overall structural view of the scene, thus helping to reasonably estimate the optical flow. The single image denoiser is built upon the residual dense blocks~\cite{zhang2018residual} due to its simple yet effective design. While we denoise the SPD captured frames, we also perform $3$D shallow-feature extraction on the Gaussian noise induced high-bit frames $\mathbf{x}_{\text{t}}$ to extract useful  spatio-temporal details. We adopt DDPM's self-conditioning mechanism~\cite{ho2020denoising} to ensure stable iterative generation.

\textbf{Flow estimation and warping for Temporal Feature Fusion.} Most existing methods in literature utilize pre-trained optical flow modules during both training and inference. However, due to low-light and noticeably different camera sensor parameters compared to that of CMOS image sensors, a learnable optical flow module that is end-to-end trained along with the other modules is preferred for optimal performance. Efficacy of this approach has already been proven in~\cite{chennuri2024quiver}. We use existing SPyNet~\cite{ranjan2017optical} architecture for the learnable flow estimation stage. Bi-directional warping and Temporal Feature Fusion is performed simultaneously using a recurrent module as shown in Figure~\ref{fig:qudi}. For every input reference frame, this framework helps QuDi extract valuable spatio-temporal information from the corresponding neighboring frames. 

\textbf{Autoencoder to produce the restored frame.} Post optical flow estimation, warping and temporal feature fusion, we have spatio-temporally feature maps extracted both from the denoised quanta frames and the high-bit Gaussian noise induced high-bit frames. As part of proving the efficacy of diffusion models in restoring noisy quanta frames, we utilize DDPM's U-Net architecture~\cite{ho2020denoising} with few modifications on the encoder side to fuse the high-bit and low-bit spatio-temporal features. Overall, the autoencoder is responsible for filtering out the signal from the noise to produce enhanced perceptual quality restored frames that are close to the actual scene being captured. 

\textbf{Quanta Forward Model to Generate QIS data.} To support better feature estimation during the reverse diffusion process, we introduce a forward model that simulates SPD measurements at each time step. While the captured low-bit data ($\mathbf{y}$) remains fixed, we use the previous estimate $\mathbf{\hat{x}}_\theta(\mathbf{x}_{\text{t}+1}, \mathbf{y})$ together with a time-step-dependent simulator to generate synthetic QIS data. This simulator produces SPD measurements with increasing average photon flux as we move through the diffusion steps, resulting in more confident and reliable measurements over time. By fusing this simulated data with the captured input, we enable the pre-denoiser, flow estimator, and temporal fusion modules to extract more stable and accurate features. Integrating the simulator into the diffusion framework also implicitly trains the model to learn the physics of the forward acquisition process.

% The denoised high-perceptual-quality frame $\mathbf{x}^\text{r}_{\text{t}-1}$ (Figure~\ref{fig:qudi}) is passed through the QIS simulator (Section~\ref{subsec:qis_sim}) to produce a quanta frame for the next iteration. The simulator uses the same camera sensor parameters as for generating input, except luminance is $10\times$ higher. Thus, the diffusion model takes ``noisy'' quanta frames as input and generates relatively ``denoised'' quanta frames while producing a high-quality restored frame.

% Integrating the simulator with diffusion implicitly trains the model to learn the physics of the forward acquisition process. During training, the restored frame is compared to the ground truth (assumed as the ideal mean of the forward process). The simulator’s randomness enables QuDi to generate plausible frames resembling scenes captured by a quanta sensor.

\textbf{Loss Function.} We train QuDi with the conventional L$2$ loss function with total variation and perceptual loss as regularizers. Mathematically, the overall loss function for training can be represented as,
\begin{align}
\label{eq:loss_func}
   \mathcal{L}_{\text{QuDi}} = \lambda_1 \cdot \mathcal{L}(\mathbf{I}^{\text{GT}}, \mathbf{x}^{\text{r}}_{\text{t}}) + &\lambda_2 \cdot \mathcal{L}(\mathbf{I}^{\text{GT}}, \mathbf{x}^{\text{d}}_{\text{t}}) \;+ \notag \\
   &\lambda_3 \cdot \text{LPIPS}(\mathbf{I}^{\text{GT}}, \mathbf{x}^{\text{r}}_{\text{t}})
\end{align}
where $\mathbf{x}^{\text{r}}_{\text{t}}$, $\mathbf{x}^{\text{d}}_{\text{t}}$ are the restored and predenoised frames, respectively. Lastly, $\mathcal{L}$ is a combination of L$2$ and total variation losses.

\section{Experiments}
\subsection{Experimental Settings}
\noindent \textbf{Baselines}. We compare the proposed method with eight existing dynamic scene reconstruction algorithms, namely Transform Denoise \cite{chan2016images}, QBP \cite{ma2020quanta}, 
Student-Teacher \cite{chi2020dynamic}, 
RVRT \cite{liangRecurrentVideoRestoration2022a}, EMVD \cite{maggioniEfficientMultiStageVideo2021}, FloRNN \cite{liUnidirectionalVideoDenoising2022}, MemDeblur \cite{jiMultiScaleMemoryBasedVideo2022}, Spk2ImgNet \cite{zhao2021spk2imgnet}, and QUIVER~\cite{chennuri2024quiver}. We also add an off-the-shelf denoiser BM$3$D \cite{dabovBM3D2007} to QBP, denoted QBP (+BM$3$D), as a baseline for comparison. As we will discuss in Section \ref{subsec:results}, QuDi beats all the baselines, both quantitatively and qualitatively. 

\noindent \textbf{Training.} We adopt the $2000$ fps high-speed dataset, I$2$-$2000$fps open-sourced by~\cite{chennuri2024quiver} and employ it as the training dataset for all the deep-learning models in our experiments. Each training sample is fetched on the fly from each clip. A training sample here is defined as a tuple containing the ground-truth/target frames and the $3$-bit quanta frames simulated at $3.25, 9.75, 26$ photons-per-pixel (PPP) ($\sim1$ lux assuming a $1.1 \mu\text{m}^2$ pixel pitch and a $1/2000$ second exposure time) using the image formation model described in Section \ref{subsec:qis_sim}.

\noindent We utilize the function mentioned in Eq.~\ref{eq:loss_func} as the cost function for training QuDi with regularization parameters $\lambda_1 = 1$, $\lambda_2 = 0.2$, and $\lambda_3 = 0.3$. The training data is extracted with a patch size of $256 \times 256$ and a batch size of $32$. The network is trained using Adam optimizer along with a cosine scheduled learning rate of $2\times10^{-5}$. QuDi takes approximately $3$ days to train on $8$ NVIDIA A$100$ GPUs using Pytorch.

\noindent \textbf{Testing}. To effectively analyze the performance of various methods, we utilize $31$ videos from I$2$-$2000$FPS containing various motion types, shapes, and speeds. Lastly, to measure the performance on real-world data, we collect and utilize the real-SPAD data using a SPAD sensor~\cite{duttonSPADBasedQVGAImage2016}. More details will be discussed in Section \ref{subsec:results}. For QuDi, with $11$ SPD frames as input, generating $11$ output frames over $10$ diffusion time-steps takes approximately $2.5$ seconds on an A100 GPU.

\subsection{Results}
\label{subsec:results}
\begin{table}[t]
    \centering
    \resizebox{1\linewidth}{!}{
    \begin{tabular}{c|cccccc} \hline
        Photons-Per-Pixel (PPP) & \multicolumn{2}{c}{$3.25$} & \multicolumn{2}{c}{$9.75$} & \multicolumn{2}{c}{$26$}\\ \hline
        Method & PSNR$\uparrow$ & SSIM$\uparrow$ & PSNR$\uparrow$ & SSIM$\uparrow$ & PSNR$\uparrow$ & SSIM$\uparrow$\\ \hline
        Transform Denoise~\cite{chan2016images} & $21.3170$ & $0.7184$ & $23.1521$ & $0.7671$ & $22.3096$ & $0.7811$\\
        QBP~\cite{ma2020quanta} & $15.9411$ & $0.1293$ & $19.1856$ & $0.2654$ & $20.7978$ & $0.4114$\\
        QBP (+ BM$3$D~\cite{dabovBM3D2007}) & $21.5476$ & $0.7033$ & $22.2001$ & $0.6899$ & $22.8617$ & $0.7832$\\
        Student-Teacher~\cite{chi2020dynamic} & $18.7200$ & $0.4006$ & $16.5195$ & $0.2479$ & $13.2889$ & $0.0735$\\
        RVRT~\cite{liangRecurrentVideoRestoration2022a} & $19.4115$ & $0.3539$ & $21.6714$ & $0.4568$ & $21.7528$ & $0.4968$\\
        EMVD~\cite{maggioniEfficientMultiStageVideo2021} & $20.0194$ & $0.5873$ & $21.0559$ & $0.6048$ & $23.4053$ & $0.5576$\\
        FloRNN~\cite{liUnidirectionalVideoDenoising2022} & $21.0341$ & $0.6785$ & $25.6132$ & $0.7091$ & $27.8520$ & $0.7784$\\
        MemDeblur~\cite{jiMultiScaleMemoryBasedVideo2022} & $19.4877$ & $0.3868$ & $14.4906$ & $0.1112$ & $16.0058$ & $0.1712$\\
        Spk2ImgNet~\cite{zhao2021spk2imgnet} & $20.3945$ & $0.5642$ & $19.6665$ & $0.6733$ & $14.9769$ & $0.6861$\\
        QUIVER~\cite{chennuri2024quiver} & \underline{26.2143} & \underline{0.7897} & \underline{26.8058} & \underline{0.8250} & \underline{27.9377} & \underline{0.8446}\\
        \hline
        \rowcolor[HTML]{C0C0C0} 
        QuDi (Ours) & \textbf{28.6408} & \textbf{0.8106} & \textbf{30.5170} & \textbf{0.8621} & \textbf{32.5538} & \textbf{0.8877}\\ \hline
    \end{tabular}
    }
\caption{Performance Comparison on the I$2$-$2000$fps dataset. QuDi out-performs all the baselines by a significant margin.}
\label{tab:syn_results}
\end{table}
\begin{figure}[ht!]
    \centering
    \includegraphics[width=0.95\linewidth]{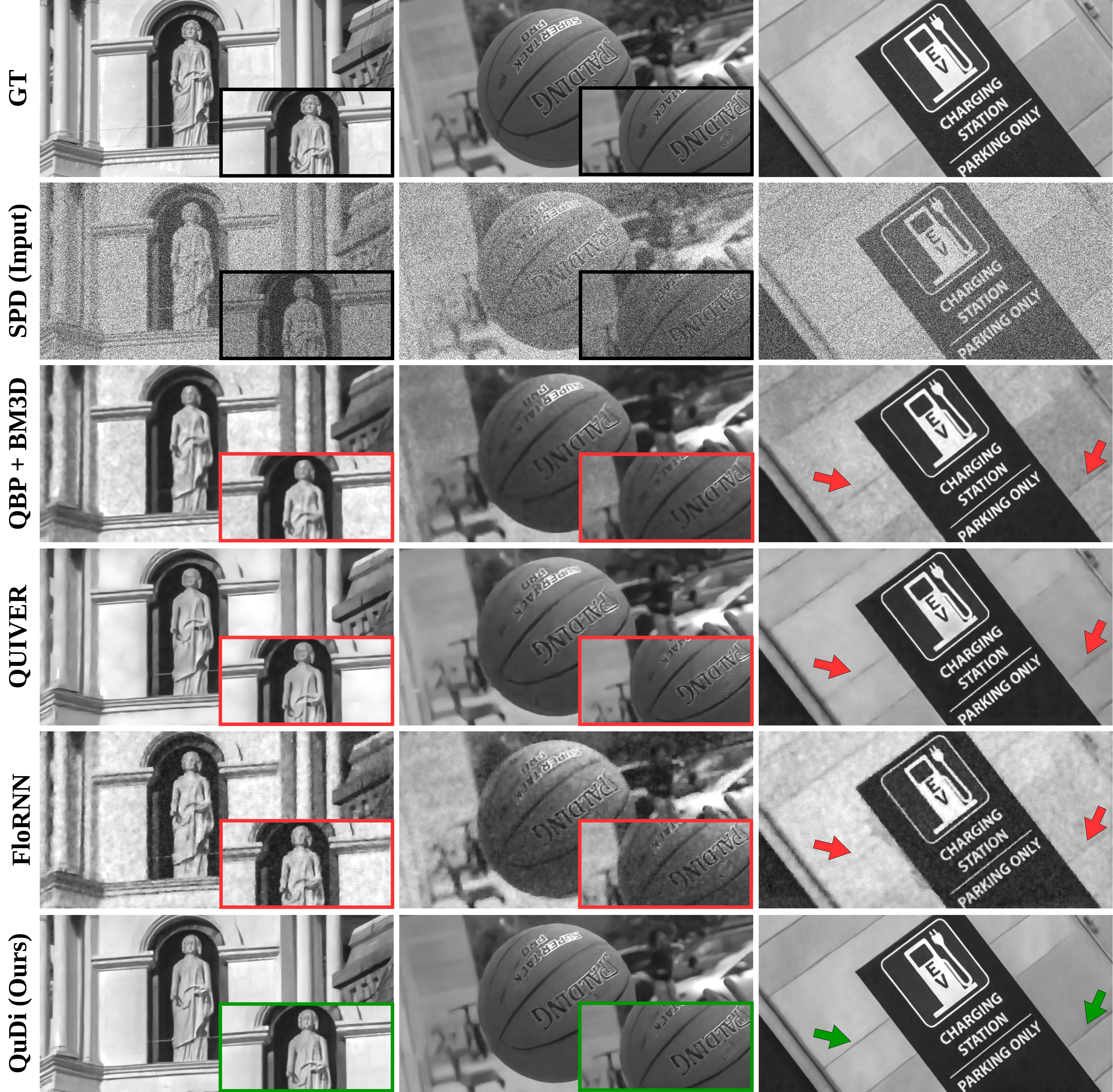}
    % \vspace{-2.5ex}
    \caption{\textbf{Synthetic Data Results}. Visual comparisons of reconstructed frames using test videos from the I$2$-$2000$FPS dataset. For a fair comparison, all methods use $11$ $3$-bit quanta frames simulated at 3.25 PPP per frame ($\sim 1$ Lux) to generate a restored frame. \textit{Best viewed in zoom}.
}
    \label{fig:syn_res}
\end{figure}
\textbf{Synthetic Data Experiments.} We evaluate synthetic data experiments using $3$-bit SPD frames generated by the quanta image simulator (Section~\ref{subsec:qis_sim}) at $3.25$, $9.75$, and $26$ Photons-Per-Pixel (PPP) per-frame as input for all methods, including QuDi. Table~\ref{tab:syn_results} presents quantitative metrics: PSNR, SSIM. Quantitative results show that QuDi outperforms the state-of-the-art QUIVER~\cite{chennuri2024quiver} and other baselines by a significant margin. 

Visual results (Figure~\ref{fig:syn_res}) demonstrate that QuDi captures high-frequency details while simultaneously managing motion and noise. High-speed reconstructed SPD video comparisons reveal that QuDi introduces fewer temporal distortions than other methods, highlighting the effectiveness of diffusion models for quanta restoration.

\begin{figure}[ht!]
    \centering
    \includegraphics[width=0.95\linewidth]{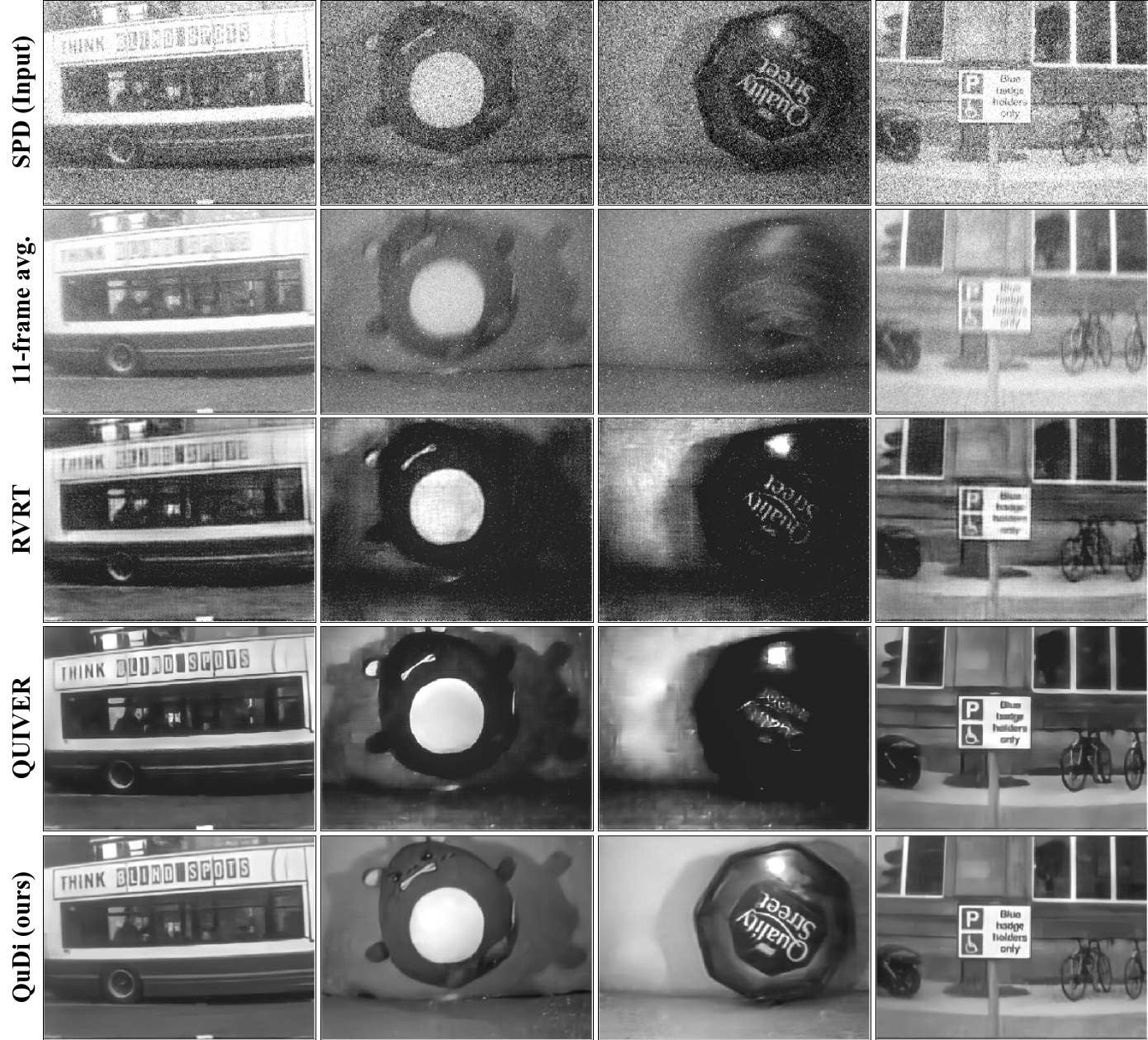}
    % \vspace{-2.5ex}
    \caption{\textbf{Performance on Real data}. Results indicate similar performance in the case of slow motion, however, QuDi outperforms QUIVER in the case of fast motion. We thank Prof. Istvan Gyongy (University of Edinburgh) for sharing data with us.}
    \label{fig:real_res}
\end{figure}
\textbf{Real Data Experiments.} Real data was collected using a SPAD sensor \cite{duttonSPADBasedQVGAImage2016}, consisting of binary frames recorded at $10000$ FPS with a spatial resolution of $240\times320$. These binary frames, due to SPADs' zero read noise, were summed to create $3$-bit frames with an average photon count of $4.9$ photons per pixel (PPP). 

Visual comparisons between RVRT~\cite{liangRecurrentVideoRestoration2022a}, QUIVER~\cite{chennuri2024quiver} and QuDi are shown in Figure~\ref{fig:real_res}. While RVRT performs the worst, for slow motion, QUIVER captures better high-frequency background details but often produces saturated outputs, whereas QuDi generates more reliable details. In fast motion scenarios, QuDi significantly outperforms QUIVER, which struggles with motion and cross-sensor adaptability. QuDi thus sets a new state-of-the-art in real-data quanta restoration across diverse scenarios.

% \textit{\textbf{Note on real-data experiments:}} SPDs, or Quanta Image Sensors (QIS), include CMOS-QIS and SPAD-QIS types. While real data is collected using an SPAD-QIS, our experiments primarily use a CMOS-QIS simulator, as modeling SPAD-QIS data acquisition is complex. QUIVER is trained with CMOS-QIS parameters equivalent to those of the SPAD-QIS used for data collection. To analyze the cross sensor performance, we utilize QuDi trained with CMOS-QIS parameters as described in Section~\ref{subsec:qis_sim}, which significantly differ from the SPAD-QIS parameters. Despite this mismatch, QuDi consistently achieves better visual performance than QUIVER.

\section{Conclusion and Future Work}
In this project, we explore the field of quanta video restoration. Existing methods like QBP~\cite{ma2020quanta}, QUIVER~\cite{chennuri2024quiver} perform well in slow motion scenarios, however produce oversmoothened or distorted outputs for fast motion cases with limited input-frames. Therefore, we explore diffusion models specifically designed to restore quanta videos for producing high-perceptual quality results while being close to the ground truth. Experiments indicate that our algorithm QuDi beats the state-of-the-art by a significant margin. 

Although we perform the best among all the methods, there are still observable temporal inconsistencies and distortions in the output videos at $\sim 1$ Lux. Therefore, the future work shall include designing quanta video consistency modules that minimize such distortions. 

\section{Acknowledgements}
This work was done as part of an internship at Dolby Laboratories. Authors thank Jonathan Miller, Ruixiang Chai, William M
Villarreal, Anustup Choudhury, and Jingxi Chen for providing valuable inputs throughout the duration of the project.
\bibliographystyle{IEEEbib}
\bibliography{refs}

\end{document}